\documentstyle[twoside,epsf]{article}

\catcode`\@=11
\long\def\@makefntext#1{
\protect\noindent \hbox to 3.2pt {\hskip-.9pt
$^{{\eightrm\@thefnmark}}$\hfil}#1\hfill}               

\def\@makefnmark{\hbox to 0pt{$^{\@thefnmark}$\hss}}    

\def\ps@myheadings{\let\@mkboth\@gobbletwo
\def\@oddhead{\hbox{}
\rightmark\hfil\eightrm\thepage}
\def\@oddfoot{}\def\@evenhead{\eightrm\thepage\hfil
\leftmark\hbox{}}\def\@evenfoot{}
\def\sectionmark##1{}\def\subsectionmark##1{}}

\oddsidemargin=\evensidemargin
\newcounter{sectionc}\newcounter{subsectionc}\newcounter{subsubsectionc}
\renewcommand{\section}[1] {\vspace{12pt}\addtocounter{sectionc}{1}
\setcounter{subsectionc}{0}\setcounter{subsubsectionc}{0}\noindent
        {\tenbf\thesectionc. #1}\par\vspace{5pt}}
\renewcommand{\subsection}[1] {\vspace{12pt}\addtocounter{subsectionc}{1}
        \setcounter{subsubsectionc}{0}\noindent
        {\bf\thesectionc.\thesubsectionc. {\kern1pt \bfit #1}}\par\vspace{5pt}}
\renewcommand{\subsubsection}[1] {\vspace{12pt}\addtocounter{subsubsectionc}{1}
        \noindent{\tenrm\thesectionc.\thesubsectionc.\thesubsubsectionc.
        {\kern1pt \tenit #1}}\par\vspace{5pt}}
\newcommand{\nonumsection}[1] {\vspace{12pt}\noindent{\tenbf #1}
        \par\vspace{5pt}}

\newcounter{appendixc}
\newcounter{subappendixc}[appendixc]
\newcounter{subsubappendixc}[subappendixc]
\renewcommand{\thesubappendixc}{\Alph{appendixc}.\arabic{subappendixc}}
\renewcommand{\thesubsubappendixc}
        {\Alph{appendixc}.\arabic{subappendixc}.\arabic{subsubappendixc}}

\renewcommand{\appendix}[1] {\vspace{12pt}
        \refstepcounter{appendixc}
        \setcounter{figure}{0}
        \setcounter{table}{0}
        \setcounter{lemma}{0}
        \setcounter{theorem}{0}
        \setcounter{corollary}{0}
        \setcounter{definition}{0}
        \setcounter{equation}{0}
        \renewcommand{\thefigure}{\Alph{appendixc}.\arabic{figure}}
        \renewcommand{\thetable}{\Alph{appendixc}.\arabic{table}}
        \renewcommand{\theappendixc}{\Alph{appendixc}}
        \renewcommand{\thelemma}{\Alph{appendixc}.\arabic{lemma}}
        \renewcommand{\thetheorem}{\Alph{appendixc}.\arabic{theorem}}
        \renewcommand{\thedefinition}{\Alph{appendixc}.\arabic{definition}}
        \renewcommand{\thecorollary}{\Alph{appendixc}.\arabic{corollary}}
        \renewcommand{\theequation}{\Alph{appendixc}.\arabic{equation}}
        \noindent{\tenbf Appendix \theappendixc #1}\par\vspace{5pt}}
\newcommand{\subappendix}[1] {\vspace{12pt}
        \refstepcounter{subappendixc}
        \noindent{\bf Appendix \thesubappendixc. {\kern1pt \bfit #1}}
        \par\vspace{5pt}}
\newcommand{\subsubappendix}[1] {\vspace{12pt}
        \refstepcounter{subsubappendixc}
        \noindent{\rm Appendix \thesubsubappendixc. {\kern1pt \tenit #1}}
        \par\vspace{5pt}}

\newcommand{\textlineskip}{\baselineskip=13pt}
\newcommand{\smalllineskip}{\baselineskip=10pt}

\def\eightcirc{
\begin{picture}(0,0)
\put(4.4,1.8){\circle{6.5}}
\end{picture}}
\def\eightcopyright{\eightcirc\kern2.7pt\hbox{\eightrm c}}

\newcommand{\copyrightheading}[1]
        {\vspace*{-2.5cm}\smalllineskip{\flushleft
{\bf OUT--4102--58\hfill hep-ph/9504400}}}

\def\abstracts#1#2#3{{
        \centering{\begin{minipage}{4.5in}\baselineskip=10pt\footnotesize
        \parindent=0pt #1\par
        \parindent=15pt #2\par
        \parindent=15pt #3
        \end{minipage}}\par}}

\newcommand{\bibit}{\nineit}
\newcommand{\bibbf}{\ninebf}
\renewenvironment{thebibliography}[1]
        {\frenchspacing
         \ninerm\baselineskip=11pt
         \begin{list}{\arabic{enumi}.}
        {\usecounter{enumi}\setlength{\parsep}{0pt}
         \setlength{\leftmargin 17pt}{\rightmargin 0pt}   
         \setlength{\itemsep}{0pt} \settowidth
        {\labelwidth}{#1.}\sloppy}}{\end{list}}

\newcounter{itemlistc}
\newcounter{romanlistc}
\newcounter{alphlistc}
\newcounter{arabiclistc}

\newcommand{\fcaption}[1]{
        \refstepcounter{figure}
        \setbox\@tempboxa = \hbox{\footnotesize Fig.~\thefigure. #1}
        \ifdim \wd\@tempboxa > 5in
           {\begin{center}
        \parbox{5in}{\footnotesize\smalllineskip Fig.~\thefigure. #1}
            \end{center}}
        \else
             {\begin{center}
             {\footnotesize Fig.~\thefigure. #1}
              \end{center}}
        \fi}

\newcommand{\tcaption}[1]{
        \refstepcounter{table}
        \setbox\@tempboxa = \hbox{\footnotesize Table~\thetable. #1}
        \ifdim \wd\@tempboxa > 5in
           {\begin{center}
        \parbox{5in}{\footnotesize\smalllineskip Table~\thetable. #1}
            \end{center}}
        \else
             {\begin{center}
             {\footnotesize Table~\thetable. #1}
              \end{center}}
        \fi}

\def\@citex[#1]#2{\if@filesw\immediate\write\@auxout
        {\string\citation{#2}}\fi
\def\@citea{}\@cite{\@for\@citeb:=#2\do
        {\@citea\def\@citea{,}\@ifundefined
        {b@\@citeb}{{\bf ?}\@warning
        {Citation `\@citeb' on page \thepage \space undefined}}
        {\csname b@\@citeb\endcsname}}}{#1}}

\newif\if@cghi
\def\cite{\@cghitrue\@ifnextchar [{\@tempswatrue
        \@citex}{\@tempswafalse\@citex[]}}
\def\citelow{\@cghifalse\@ifnextchar [{\@tempswatrue
        \@citex}{\@tempswafalse\@citex[]}}
\def\@cite#1#2{{$\null^{#1}$\if@tempswa\typeout
        {IJCGA warning: optional citation argument
        ignored: `#2'} \fi}}

\def\pmb#1{\setbox0=\hbox{#1}
        \kern-.025em\copy0\kern-\wd0
        \kern.05em\copy0\kern-\wd0
        \kern-.025em\raise.0433em\box0}

\def\fnt#1#2{\footnotetext{\kern-.3em
        {$^{\mbox{\scriptsize #1}}$}{#2}}}

\def\fpage#1{\begingroup
\voffset=.3in
\thispagestyle{empty}\begin{table}[b]\centerline{\footnotesize #1}
        \end{table}\endgroup}

\def\runninghead#1#2{\pagestyle{myheadings}
\markboth{{\protect\footnotesize\it{\quad #1}}\hfill}
{\hfill{\protect\footnotesize\it{#2\quad}}}}
\font\tenrm=cmr10
\font\tenit=cmti10
\font\tenbf=cmbx10
\font\bfit=cmbxti10 at 10pt
\font\ninerm=cmr9
\font\nineit=cmti9
\font\ninebf=cmbx9
\font\eightrm=cmr8

\def\qed{\hbox{${\vcenter{\vbox{                        
   \hrule height 0.4pt\hbox{\vrule width 0.4pt height 6pt
   \kern5pt\vrule width 0.4pt}\hrule height 0.4pt}}}$}}

\def\bsc{{\sc a\kern-6.4pt\sc a\kern-6.4pt\sc a}}       
\def\bflatex{\bf L\kern-.30em\raise.3ex\hbox{\bsc}\kern-.14em
T\kern-.1667em\lower.7ex\hbox{E}\kern-.125em X}

\begin{document}

\runninghead{Multiloop Calculations in Heavy Quark Effective Theory}
{Multiloop Calculations in Heavy Quark Effective Theory}

\normalsize\textlineskip
\thispagestyle{empty}
\setcounter{page}{1}

\copyrightheading{}                     

\vspace*{0.88truein}

\fpage{1}
\centerline{\bf Multiloop Calculations in Heavy Quark Effective Theory%
\footnote{
Talk presented by A.~G.~Grozin at the AIHENP--95 workshop, Pisa, April 1995}}
\vspace*{0.37truein}
\centerline{\footnotesize D.~J.~BROADHURST\footnote{
D.Broadhurst@open.ac.uk}
\hspace{3pt} and \hspace{3pt}A.~G.~GROZIN\footnote{
A.Grozin@open.ac.uk;\\
\hspace*{3pt}on leave of absence from Budker Institute
of Nuclear Physics, Novosibirsk, Russia}}
\vspace*{0.015truein}
\centerline{\footnotesize\it Physics Department, Open University}
\baselineskip=10pt
\centerline{\footnotesize\it Milton Keynes, MK7 6AA, UK}
\vspace*{0.225truein}
\abstracts{We review algorithmic methods for two--loop calculations in HQET,
and the analogous methods for on--shell QCD,
needed for matching HQET to QCD.}{}{}
\vspace*{1pt}\textlineskip

\section{Introduction}
\vspace*{-0.5pt}\noindent
In this talk, we review methods for two--loop calculations
in heavy quark effective theory\cite{HQET}.
In HQET, the heavy--quark propagator is $S(k)=1/(k\cdot v+i0)$,
and the quark--gluon vertex is ${\rm i}g t^av_\mu$,
with a heavy--quark velocity
$v=(1,{\bf 0})$, in the rest frame.
We shall discuss algorithmic methods,
suitable for computer--algebra implementation.
All of them are based on dimensional regularization,
with space--time dimension $d=4-2\varepsilon$,
and integration by parts\cite{CT}, a powerful calculational method
based on the simple observation that integrals of full derivatives vanish.
Tensor integrals are handled by decomposition into
a suitable tensor basis,
with coefficients that can be expressed in terms of scalar integrals,
by solving linear systems of equations.
In all cases which we shall discuss,
numerators can be expressed in terms of factors of the denominator.

In Section~2, we recall the well--known integration--by--parts
algorithm for calculation of massless two--point integrals\cite{CT}.
This example is used to introduce notations
and to set up the pattern which will be repeated in other cases.
In Section~3, we discuss the corresponding method in HQET\cite{BG1}.
This method was generalized\cite{BBG} to HQET three--point integrals,
expanded in the velocity change $v'-v$, to any finite order, as discussed
in Section~4.
In order to obtain QCD/HQET matching conditions\cite{BG3},
it is necessary to equate QCD and HQET on--shell matrix elements.
The latter are trivial (apart from massive quark loops).
The former require methods for on--shell
massive QCD calculations\cite{GBGS,BGS,B,FT}, discussed in
Section~5.

\section{Massless two--point integrals}
\begin{figure}[ht]
\epsffile{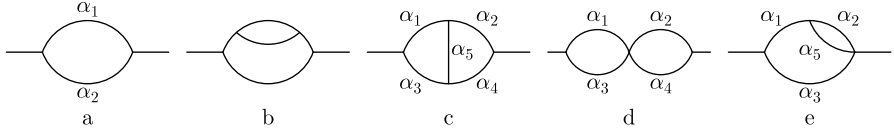}
\caption{Massless two--point diagrams}
\label{FM2}
\end{figure}

\vspace*{-0.5pt}\noindent
We write the massless one--loop integral of Fig.~\ref{FM2}a,
with arbitrary indices, as
\begin{equation}
\int\frac{{\rm d}^d k}{(2\pi)^d}\frac{1}{(-k^2)^{\alpha_1}
(-(k+p)^2)^{\alpha_2}}
=\frac{{\rm i}
(-p^2)^{d/2-\alpha_1-\alpha_2}}{(4\pi)^{d/2}}G(\alpha_1,\alpha_2).
\end{equation}
It can be calculated by using the Feynman parameterization
\begin{equation}
\frac{1}{a^\alpha b^\beta}
=\frac{\Gamma(\alpha+\beta)}{\Gamma(\alpha)\Gamma(\beta)}\int\limits_0^1
\frac{x^{\alpha-1}(1-x)^{\beta-1}{\rm d}x}{[ax+b(1-x)]^{\alpha+\beta}}\,
\label{Feyn}
\end{equation}
or by using Fourier transformation to coordinate space and back.
The result is
\begin{equation}
G(\alpha_1,\alpha_2)
=\frac{\Gamma(\alpha_1+\alpha_2-d/2)\Gamma(d/2-\alpha_1)\Gamma(d/2-\alpha_2)}
{\Gamma(\alpha_1)\Gamma(\alpha_2)\Gamma(d-\alpha_1-\alpha_2)}.
\label{Gres}
\end{equation}
All one--loop integrals with integer indices can therefore be expressed
in terms of
$G_0=\Gamma(1+\varepsilon)\Gamma^2(1-\varepsilon)/\Gamma(1-2\varepsilon)$,
with rational coefficients.

Two--loop integrals of the form of Fig.~\ref{FM2}b are evaluated
by repeated use of~(\ref{Gres}).
The integral of Fig.~\ref{FM2}c is non--trivial. We write it as
\begin{eqnarray}
&&\int\frac{{\rm d}^d k}{(2\pi)^d}\frac{{\rm d}^d l}{(2\pi)^d}\frac{1}
{(-k^2)^{\alpha_1}(-l^2)^{\alpha_2}(-(k+p)^2)^{\alpha_3}(-(l+p)^2)^{\alpha_4}
(-(k-l)^2)^{\alpha_5}}
\nonumber\\
&&=-\frac{(-p^2)^{d-\sum\alpha}}{(4\pi)^d}
F(\alpha_1,\alpha_2,\alpha_3,\alpha_4,\alpha_5).
\label{Fdef}
\end{eqnarray}
Integration by parts gives the recurrence relation
\begin{equation}
(d-\alpha_1-\alpha_3-2\alpha_5)F
=\left[\alpha_1{\bf1^+}({\bf5^-}-{\bf2^-})+\alpha_3{\bf3^+}({\bf5^-}
-{\bf4^-})\right]F,
\label{Triangle}
\end{equation}
where, for example, ${\bf1^+}$ increases $\alpha_1$,
and ${\bf2^-}$ decreases $\alpha_2$.
Applying this relation sufficiently many times,
we can kill one of the lines 2, 4, or 5, in Fig.~\ref{FM2}c.
The resulting integrals, in Figs.~\ref{FM2}d,e, are trivial:
\begin{eqnarray}
&&F(\alpha_1,\alpha_2,\alpha_3,\alpha_4,0)
=G(\alpha_1,\alpha_3)G(\alpha_2,\alpha_4),
\nonumber\\
&&F(\alpha_1,\alpha_2,\alpha_3,0,\alpha_5)
=G(\alpha_1+\alpha_2+\alpha_5-d/2,\alpha_3)G(\alpha_2,\alpha_5).
\label{Ftriv}
\end{eqnarray}
All two--loop integrals can be thus expressed in terms two structures:
$G_0^2$ and
$G_1=\Gamma(1+2\varepsilon)\Gamma^3(1-\varepsilon)/\Gamma(1-3\varepsilon)$,
with rational coefficients.
This algorithm\cite{CT} is implemented in REDUCE\cite{ST,BG1},
and as part of the 3--loop package MINCER\cite{GLST}.

\section{HQET two--point integrals}
\begin{figure}[ht]
\epsffile{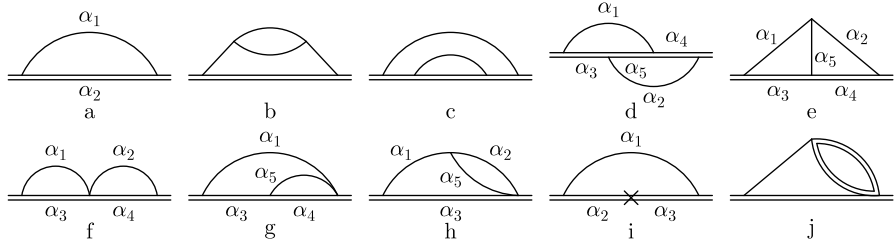}
\caption{Two--point diagrams in HQET or on--shell QCD}
\label{FH2}
\end{figure}

\vspace*{-0.5pt}\noindent
We write the HQET one--loop integral of Fig.~\ref{FH2}a,
with arbitrary indices, as
\begin{equation}
\int\frac{{\rm d}^d k}{(2\pi)^d}\frac{1}{(-k^2)^{\alpha_1}}
\left(\frac{\omega}{k_0+\omega}\right)^{\alpha_2}
=\frac{{\rm i}(-2\omega)^{d-2\alpha_1}}{(4\pi)^{d/2}}I(\alpha_1,\alpha_2).
\label{Idef}
\end{equation}
It can be calculated, using the parameterization
\begin{equation}
\frac{1}{a^\alpha b^\beta}
=\frac{\Gamma(\alpha+\beta)}{\Gamma(\alpha)\Gamma(\beta)}\int\limits_0^\infty
\frac{y^{\beta-1}{\rm d}y}{(a+by)^{\alpha+\beta}},
\label{FeynH}
\end{equation}
where $y$ has dimensions of energy. The result is
\begin{equation}
I(\alpha_1,\alpha_2)
=\frac{\Gamma(2\alpha_1+\alpha_2-d)\Gamma(d/2-\alpha_1)}
{\Gamma(\alpha_1)\Gamma(\alpha_2)}.
\label{Ires}
\end{equation}
All one--loop integrals with integer indices can be expressed in terms of
$I_0=\Gamma(1+2\varepsilon)\Gamma(1-\varepsilon)$
with rational coefficients.

The two--loop integrals of Figs.~\ref{FH2}b,c are trivial.
Multiplying the integrand of Fig.~\ref{FH2}d by
\begin{equation}
1=\frac{1}{\omega}\left[(k_0+\omega)+(l_0+\omega)-(k_0+l_0+\omega)\right],
\label{Id}
\end{equation}
we can kill one of the heavy lines,
reducing this integral to trivial ones, like those of Figs.~\ref{FH2}f,g.
Only the integral of Fig.~\ref{FH2}e is non--trivial. We write it as
\begin{eqnarray}
&&\int\frac{{\rm d}^d k}{(2\pi)^d}\frac{{\rm d}^d l}{(2\pi)^d}\frac{1}
{(-k^2)^{\alpha_1}(-l^2)^{\alpha_2}(-(k-l)^2)^{\alpha_5}}
\left(\frac{\omega}{k_0+\omega}\right)^{\alpha_3}
\left(\frac{\omega}{l_0+\omega}\right)^{\alpha_4}
\nonumber\\
&&=-\frac{(-2\omega)^{2(d-\alpha_1-\alpha_2-\alpha_5)}}{(4\pi)^d}
I(\alpha_1,\alpha_2,\alpha_3,\alpha_4,\alpha_5).
\label{I5def}
\end{eqnarray}
The recurrence relation
\begin{eqnarray}
&&(d-\alpha_1-\alpha_3-\alpha_4-2\alpha_5+1)I
\nonumber\\
&&=\left[\alpha_1{\bf1^+}({\bf5^-}-{\bf2^-})
+(2(d-\alpha_1-\alpha_2-\alpha_5)-\alpha_3-\alpha_4+1){\bf4^-}\right]I
\label{TriangleH}
\end{eqnarray}
allows us to kill one of the lines 2, 4, or 5, in Fig.~\ref{FH2}e.
The resulting integrals, in Figs.~\ref{FH2}f,g,h, are trivial:
\begin{eqnarray}
&&I(\alpha_1,\alpha_2,\alpha_3,\alpha_4,0)
=I(\alpha_1,\alpha_3)I(\alpha_2,\alpha_4),
\nonumber\\
&&I(\alpha_1,0,\alpha_3,\alpha_4,\alpha_5)
=I(\alpha_1,\alpha_3+\alpha_4+2\alpha_5-d)I(\alpha_5,\alpha_4),
\label{Itriv}\\
&&I(\alpha_1,\alpha_2,\alpha_3,0,\alpha_5)
=I(\alpha_1+\alpha_2+\alpha_5-d/2,\alpha_3)G(\alpha_2,\alpha_5).
\nonumber
\end{eqnarray}
All two--loop integrals can thus be expressed in terms of $I_0^2$ and
$I_1=\Gamma(1+4\varepsilon)\Gamma^2(1-\varepsilon)$,
with rational coefficients.

This algorithm was proposed in\cite{BG1}, where it was
implemented in REDUCE.
It was used for calculation of:
the heavy--field anomalous dimension\cite{BG1} (previously considered
in Wilson--line\cite{KS}
and on--shell\cite{BGS} QCD calculations);
the heavy--light current anomalous dimension\cite{BG1}
(also obtained using Feynman parameterization\cite{JM}
and Gegenbauer polynomial expansion,
with a more difficult momentum routing\cite{G});
and the correlator of two heavy--light currents\cite{BG2,BBBG}.
The coefficient of $\alpha_s/\pi$ in this correlator is of order 10,
making sum rules for heavy--light mesons unreliable,
in our opinion\cite{BG2}.

\section{HQET three--point integrals expanded in $v'-v$}
\vspace*{-0.5pt}\noindent
Here we consider three--point diagrams,
containing HQET propagators with two velocities, $v$
and $v'$, and depending on two residual energies, $\omega$ and $\omega'$.
Their expansions in $v'-v$ to any finite order can be evaluated algebraically.
The heavy--quark propagators are expanded in $v'-v$,
and the resulting tensor integrals are reduced to scalar integrals.
The one--loop integral of Fig.~\ref{FH2}i,
with arbitrary indices specifying the
dependence on $\omega$ and $\omega'$, is
\begin{eqnarray}
&&\!\!\!\!\int\frac{{\rm d}^d k}{(2\pi)^d}\frac{1}{(-k^2)^{\alpha_1}}
\left(\frac{\omega}{k_0+\omega}\right)^{\alpha_2}
\left(\frac{\omega'}{k_0+\omega'}\right)^{\alpha_3}
=\frac{{\rm i}(-2\omega')^{d-2\alpha_1}}{(4\pi)^{d/2}}
I\left(\alpha_1,\alpha_2,\alpha_3;\frac{\omega}{\omega'}\right),
\nonumber\\
&&\!\!\!\!I(\alpha_1,\alpha_2,\alpha_3;x)
=I(\alpha_1,\alpha_2+\alpha_3)x^{\alpha_2}
\;_2F_1(2\alpha_1+\alpha_2+\alpha_3-d,\alpha_2;\alpha_2+\alpha_3;1-x).
\nonumber
\end{eqnarray}

We write the two--loop integral of Fig.~\ref{FH2}e as
\begin{eqnarray}
&&\int\frac{{\rm d}^d k}{(2\pi)^d}\frac{{\rm d}^d l}{(2\pi)^d}
\frac{1}{(-k^2)^{\alpha_1}(-l^2)^{\alpha_2}(-(k-l)^2)^{\alpha_5}}
\left(\frac{\omega}{k_0+\omega}\right)^{\alpha_3}
\left(\frac{\omega'}{l_0+\omega'}\right)^{\alpha_4}
\nonumber\\
&&=-\frac{(-2\omega')^{2(d-\alpha_1-\alpha_2-\alpha_5)}}{(4\pi)^d}
I\left(\alpha_1,\alpha_2,\alpha_3,\alpha_4,\alpha_5;
\frac{\omega}{\omega'}\right).
\label{I2}
\end{eqnarray}
In the Feynman gauge, all diagrams generate scalar integrals with
$\alpha_{1,2,5}\le1$.
Moreover, the cases with $\alpha_{3,4}>1$
can be obtained from $I(\alpha_1,\alpha_2,1,1,\alpha_5;x)$
by differentiation with respect to $\omega$ and $\omega'$.
Then, either some of the lines are already absent,
or one of the lines 1, 2, 4, 5 can be killed by applying
the recurrence relation\cite{BBG}
\begin{eqnarray}
&&\left[\alpha_5-\alpha_1
+\frac{\omega'}{\omega}(2\alpha_1+\alpha_5+\alpha_3-d)\right]I
\nonumber\\
&&=\left[\alpha_5\left(1-\frac{\omega'}{\omega}\right){\bf5^+}
({\bf1^-}-{\bf2^-})
+\alpha_1{\bf1^+}({\bf2^-}-{\bf5^-})
+\frac{\omega'}{\omega}\alpha_3{\bf3^+}{\bf4^-}\right]I.
\label{Triangle3}
\end{eqnarray}
This produces the trivial integrals of Figs.~\ref{FH2}f,h,
and the integral of Fig.~\ref{FH2}g, given by $I(\alpha_5,\alpha_4)
I\left(\alpha_1,\alpha_3,\alpha_4+2\alpha_5-d;\frac{\omega}{\omega'}\right)$.

This algorithm was proposed in\cite{BBG};
we are not aware of a package that implements it.
The method was applied for checking\cite{BG}, at small velocity
transfer,
the general result
for the heavy--heavy current anomalous dimension\cite{KR},
and for obtaining\cite{BBG} the correlator of two heavy--light currents
and one heavy--heavy current, at small $v'-v$
(subsequently obtained, for all velocity transfers\cite{N},
using methods based on differential equations).
Again there are large radiative corrections. However, these
are largely cancelled in the ratio of
the three--point and two--point sum rules,
making predictions for the Isgur--Wise function\cite{BBG,N} relatively safe.

\section{On--shell massive two--point integrals}
\vspace*{-0.5pt}\noindent
Expansions of massive two--point diagrams to any finite order in $p^2-m^2$
can be evaluated algebraically.
Such on--shell diagrams are necessary for QCD/HQET matching.
We write the one--loop integral of Fig.~\ref{FH2}a,
with arbitrary indices, as
\begin{equation}
\int\frac{{\rm d}^d k}{(2\pi)^d}\frac{1}{(-k^2)^{\alpha_1}
(-(k+m v)^2+m^2)^{\alpha_2}}
=\frac{{\rm
i}m^{d-2(\alpha_1+\alpha_2)}}{(4\pi)^{d/2}}M(\alpha_1,\alpha_2).
\label{Mdef}
\end{equation}
It can be evaluated by Feynman parameterization.
Alternatively, we can use the inversion $k_\mu\to k_\mu/k^2$, after
transforming to a dimensionless euclidean loop momentum.
This transforms the on--shell massive denominator into an HQET
denominator,
and we obtain
\begin{equation}
M(\alpha_1,\alpha_2)=I(d-\alpha_1-\alpha_2,\alpha_2)
=\frac{\Gamma(\alpha_1+\alpha_2-d/2)\Gamma(d-2\alpha_1-\alpha_2)}
{\Gamma(\alpha_2)\Gamma(d-\alpha_1-\alpha_2)}.
\label{Mres}
\end{equation}
All one--loop integrals with integer indices can be expressed in terms of
$M_0=\Gamma(1+\varepsilon)$ with rational coefficients.

There are two distinct non--trivial topologies for two--loop integrals.
We shall start with the simpler type $M$, of Fig.~\ref{FH2}e,
\begin{eqnarray}
&&\int\frac{{\rm d}^d k}{(2\pi)^d}\frac{{\rm d}^d l}{(2\pi)^d}
\frac{1}{(-k^2)^{\alpha_1}(-l^2)^{\alpha_2}(-(k-l)^2)^{\alpha_5}
(-(k+m v)^2+m^2)^{\alpha_3}}
\nonumber\\
&&{}\times\frac{1}{(-(l+m v)^2+m^2)^{\alpha_4}}=
-\frac{m^{2(d-\sum\alpha)}}{(4\pi)^d}
M(\alpha_1,\alpha_2,\alpha_3,\alpha_4,\alpha_5).
\label{M5def}
\end{eqnarray}
Using inversion, we can relate it to an HQET two--loop integral
\begin{equation}
M(\alpha_1,\alpha_2,\alpha_3,\alpha_4,\alpha_5)
=I(d-\alpha_1-\alpha_3-\alpha_5,d-\alpha_2-\alpha_4-\alpha_5,
\alpha_3,\alpha_4,\alpha_5).
\label{Inv}
\end{equation}
However, this HQET integral
contains two non--integer indices, and is hence more
difficult to calculate than before.
The recurrence relation for $M$ has a form identical
to~(\ref{Triangle}),
allowing us to kill one of the lines 2, 4, or 5 of Fig.~\ref{FH2}e.
The integrals of Figs.~\ref{FH2}f,h are trivial, as in~(\ref{Itriv}),
but that of Fig.~\ref{FH2}g is not.
Special combinations of recurrence relations allow us to kill one more line.
As a result, all two--loop integrals of the type $M$ can be expressed
in terms of $M_0^2$ and $M_1=\Gamma(1+\varepsilon)\Gamma^2(1-\varepsilon)
\Gamma(1+2\varepsilon)\Gamma(1-4\varepsilon)/(\Gamma(1-2\varepsilon)
\Gamma(1-3\varepsilon))$ with rational coefficients.

We write the most difficult integrals, of the type $N$ of Fig.~\ref{FH2}d,
as
\begin{eqnarray}
&&\int\frac{{\rm d}^d k}{(2\pi)^d}\frac{{\rm d}^d l}{(2\pi)^d}
\frac{1}{(-k^2)^{\alpha_1}(-l^2)^{\alpha_2}
(-(k+m v)^2+m^2)^{\alpha_3}(-(l+m v)^2+m^2)^{\alpha_4}}
\nonumber\\
&&{}\times\frac{1}{(-(k+l+m v)^2+m^2)^{\alpha_5}}=
-\frac{m^{2(d-\sum\alpha)}}{(4\pi)^d}
N(\alpha_1,\alpha_2,\alpha_3,\alpha_4,\alpha_5).
\label{Ndef}
\end{eqnarray}
The recurrence relation
\begin{equation}
(d-2\alpha_2-\alpha_4-\alpha_5)N
=[\alpha_4{\bf4^+}{\bf2^-}+\alpha_5{\bf5^+}({\bf2^-}-{\bf3^-})]N
\label{TriangleN}
\end{equation}
allows us to kill one of the lines 2 or 3 of Fig.~\ref{FH2}d.
This gives integrals shown in Fig.~\ref{FH2}g, belonging to the class $M$,
and integrals shown in Fig.~\ref{FH2}j, which are still non--trivial.
Special combinations of recurrence relations allow us to kill one more line.
As a result, all two--loop integrals of the type $N$ can be expressed
in terms of $M_0^2$, $M_1$, and $N(0,0,1,1,1)$ with rational coefficients.
Instead of using $N(0,0,1,1,1)$ as a basis integral, it is
more convenient to use the convergent integral\cite{B0,B}
\begin{equation}
N(1,1,1,1,1)
=\pi^2\log2-{\textstyle\frac{3}{2}}\zeta(3)+O(\varepsilon),
\label{Ie}
\end{equation}
whose $O(\varepsilon)$ and $O(\varepsilon^2)$ terms have been found
by algebraic methods\cite{B}, but
are not necessary for two--loop calculations.

This algorithm is implemented in the REDUCE package RECURSOR\cite{B}
and the FORM package SHELL2\cite{FT}.
It was used for calculating heavy--quark on--shell mass\cite{GBGS}
and wave--function\cite{BGS} renormalizations,
and the QCD/HQET matching of heavy--light currents\cite{BG3}.
Coefficients of $(\alpha_s/\pi)^2$ in the matching coefficients
are typically of order 10, leading to substantial two--loop corrections
in ratios such as $f_{B^*}/f_B$.

\nonumsection{Acknowledgements}
\vspace*{-0.5pt}\noindent
We are grateful to the Royal Society and PPARC, for grants which made
our collaboration possible, to HUCAM, for travel funds,
and to P.~Gosdzinsky, for useful comments
about the three--point case\cite{BBG}.

\nonumsection{References}

\end{document}